# Super-High-Frequency Low-Loss Sezawa Mode SAW Devices in a GaN/SiC Platform

Imtiaz Ahmed, *Graduate Student Member, IEEE*, Udit Rawat, *Graduate Student Member, IEEE*, Jr-Tai Chen, and Dana Weinstein, *Senior Member, IEEE*

*Abstract*— This paper presents a comprehensive study of the performance of Sezawa surface acoustic wave (SAW) devices in SweGaN QuanFINE® ultrathin GaN/SiC platform, reaching frequencies above 14 GHz for the first time. Sezawa mode frequency scaling is achieved due to the elimination of the thick buffer layer typically present in epitaxial GaN technology. Finite element analysis (FEA) is first performed to find the range of frequencies over which the Sezawa mode is supported in the grown structure. Transmission lines and resonance cavities driven with Interdigital Transducers (IDTs) are designed, fabricated, and characterized. Modified Mason circuit models are developed for each class of devices to extract critical performance metrics. We observe a strong correlation between measured and simulated dispersion of the phase velocity ($v_p$) and piezoelectric coupling coefficient ($k^2$). Maximum $k^2$ of 0.61% and frequency-quality factor product ($f \cdot Q_m$) of $6 \times 10^{12}$ s$^{-1}$ are achieved for Sezawa resonators at 11 GHz, with a minimum propagation loss of 0.26 dB/λ for the two-port devices. Sezawa modes are observed at frequencies spanning up to 14.3 GHz, achieving a record high in GaN microelectromechanical systems (MEMS) to the best of the authors' knowledge.

## I. INTRODUCTION

SAW devices are essential components for filters, oscillators, and RF signal processing blocks in wireless communication due to their advantages of lithographically defined resonance frequency, simple fabrication process despite the use of materials that have historically been difficult to etch, low manufacturing cost, and low sensitivity to acceleration [1]–[3]. They are also widely used in sensor applications due to SAW sensitivity to various environmental factors including temperature, pressure, viscosity, humidity, etc. at the surface of the device [3]–[6]. However, a technology gap in growth and material quality has thus far limited the scaling of SAW frequencies with low losses required for Super High Frequency (SHF) MEMS components for 5G/6G and the Internet of Things (IoT). Frequency scaling is also beneficial for SAW sensors as their sensitivity increases with the resonance frequency [7]–[9].

Sezawa acoustic waves in a technology platform such as GaN on SiC provide a much-needed solution due to high phase velocity to scale to higher frequencies for given lithographic feature sizes, with the added benefits of high piezoelectric coupling and low viscoelastic losses [10], [11]. This mode exhibits improved confinement within the piezoelectric layer relative to the Rayleigh mode due to a mismatch in the acoustic impedance between the GaN epilayer and the SiC substrate, which reduces leakage through the substrate [3], [12], [13]. In contrast with suspended acoustic structures, Sezawa mode devices are solidly mounted to the bulk substrate, in this case SiC, which provides excellent thermal conductivity improving power handling and frequency stability [14]. The ability to withstand extreme temperatures, radiation tolerance, and chemical inertness makes GaN devices ideal for the harsh environment required in space exploration, military, and industrial applications [15]–[17]. Integration of electromechanical components with GaN high-speed active devices utilizing inherent 2D electron gas (2DEG) in the heterostructures in standard monolithic microwave integrated circuit (MMIC) technology reduces system-level parasitics, size, weight, cost and increases performance [18], [19]. Acoustic wave interaction with the 2DEG in GaN heterostructures known as acoustoelectric (AE) effect enables nonreciprocal amplification and velocity tuning of acoustic waves in correlators and other RF signal processing devices [20]–[22].

GaN heterostructures which are generally available commercially, are typically grown epitaxially on substrates including (111) Si, SiC, and Sapphire using a thick buffer layer composed of alternating GaN/AlN layers to minimize residual stress generated in high-temperature growth [23]. The thicker GaN layer with high dislocation and defect density in the buffer has thus far limited SAW devices to moderate frequencies. Until now, SAW devices were reported below 10 GHz, limited by lithography, acoustic velocity, and piezoelectric coupling dependence on film thickness [12], [13], [24]. In addition, SAW propagation loss and dispersion were limited by defect scattering and interface losses in the multilayer buffer. SweGaN's buffer-free QuanFINE® heterostructure [25] provides an ultrathin GaN epilayer grown on SiC which, along with electron beam (e-beam) lithography, facilitates aggressive frequency scaling. The authors previously reported on the feasibility of the QuanFINE® platform for SAW devices up to

Imtiaz Ahmed is with the Elmore Family School of Electrical and Computer Engineering, Purdue University, West Lafayette, IN 47906, USA; on leave from the Department of Electrical and Electronic Engineering, Bangladesh University of Engineering and Technology, Dhaka 1205, Bangladesh (e-mail: ahmed111@purdue.edu).
Udit Rawat and Dana Weinstein are with the Elmore Family School of Electrical and Computer Engineering, Purdue University, West Lafayette, IN 47906, USA (e-mail: rawatu@purdue.edu; danaw@purdue.edu).
Jr-Tai Chen is with SweGaN AB, Linköping, Sweden (e-mail: jr-tai.chen@swegan.se).

14 GHz [26]. This paper expands on those results with a comprehensive analysis of the dispersion of the Sezawa mode for different SAW designs, including an assessment of several key parameters such as phase velocity, coupling coefficient, propagation loss, and frequency-quality factor product.

## II. QUANFINE® GAN-ON-SIC PLATFORM

Sezawa mode devices are designed in SweGaN's QuanFINE® platform which consists of a heterostructure of AlGaN/GaN with a low defect density unintentionally doped (UID) thin GaN channel layer (Fig. 1(a)) [25]. An ultrathin AlN nucleation layer (NL) is first grown on semi-insulating 4H-SiC to initiate the high-quality growth of the structure in a metal organic chemical vapor deposition (MOCVD) reactor [27]–[29]. The QuanFINE® process does not require a thick, doped buffer layer typical in conventional GaN MMICs. A high-quality in-situ SiN passivation layer is introduced on the AlGaN barrier layer to protect the sensitive surface from external damage. High-resolution transmission electron micrographs (TEMs) of the GaN/AlN and AlN/SiC interfaces (Fig. 1(b)) demonstrate grain-free boundaries with low void and dislocation density that are essential for high-performance MEMS and MMIC devices [27]. The AlN nucleation layer significantly improves heat dissipation from the GaN channel to the high-thermal-conductivity SiC substrate due to its outstanding crystalline quality and ultra-low thermal boundary resistance (TBR), making the structure ideal for high power operation [25]. X-ray diffraction rocking curves (XRCs) exhibit full width at half maximum (FWHM) of 86 arcsec and 272 arcsec for the GaN (002) and GaN (102) reflections (Fig. 1(c)). These measurements correspond to a reduction in defect density of two orders of magnitude when compared with typical GaN epilayers of similar thickness [27], [30], enabling low viscoelastic losses for GaN electromechanical devices [20].

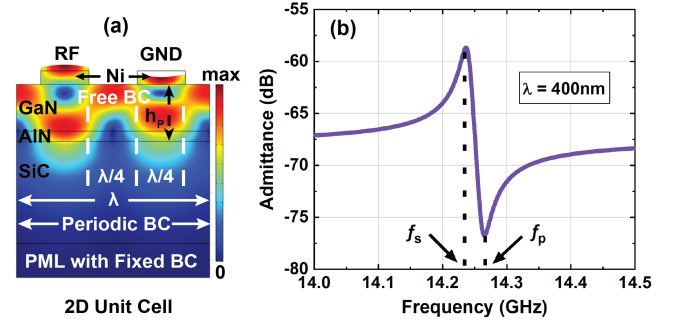

Fig. 2. 2D unit cell FEA simulation in COMSOL Multiphysics depicting (a) total displacement of Sezawa mode and (b) one-port admittance plot ($Y_{11}$) for a unit cell width ($\lambda$) of 400 nm. Series ($f_s$) and parallel ($f_p$) resonances are indicated in the plot.

## III. FEA SIMULATION

The presence of SHF Sezawa modes in the QuanFINE® heterostructure is confirmed through 2D FEA simulation using COMSOL Multiphysics. Eigenmode analysis is performed on a unit cell spanning one acoustic wavelength ($\lambda$) and IDT width ($W_{IDT}$) and gap each spanning one quarter wavelength ($\lambda/4$). The 2DEG formed in the heterostructure can be used to block electromechanical actuation by shielding the electric fields from penetrating inside the piezoelectric layers [31]. For this fundamental study of mode propagation, the 2DEG along with the AlGaN barrier and SiN passivation layers in the QuanFINE® structure are therefore removed during the fabrication process and are not included in the unit cell simulation. A pair of Ni IDT fingers with thickness of 80 nm is included to capture the effects of mass loading and interface on the resonance mode [32], [33]. Boundary Conditions (BCs) are defined as free on the top of the unit cell geometry, and periodic on both sides of the unit cell. A perfectly matched layer (PML) is introduced at the bottom of the unit cell with fixed BC to simulate radiative losses into the thick SiC substrate. A contour plot of total displacement shows the Sezawa mode in the structure at resonance (Fig. 2(a)). Simulated one-port admittance response ($Y_{11}$) between signal and ground terminals for a wavelength of 400 nm exhibits resonance ($f_s$) and anti-resonance ($f_p$) frequencies at 14.23 GHz and 14.26 GHz, respectively (Fig. 2(b)).

## IV. FABRICATION AND CHARACTERIZATION

SAW devices are fabricated on a QuanFINE® substrate using a two-mask process. We begin with a shallow blanket etch of in-situ SiN and AlGaN layers using $CHF_3/O_2$ and $BCl_3/Cl_2$ plasma inside an inductively coupled plasma–reactive ion etcher (ICP-RIE) (Fig. 3(a)). A 10-20 nm etch of the GaN channel is performed to ensure complete removal of the 2DEG heterojunction. E-beam lithography is used to pattern the IDTs followed by evaporation of Ti(5nm)/Ni(75nm) (Fig. 3(b)).

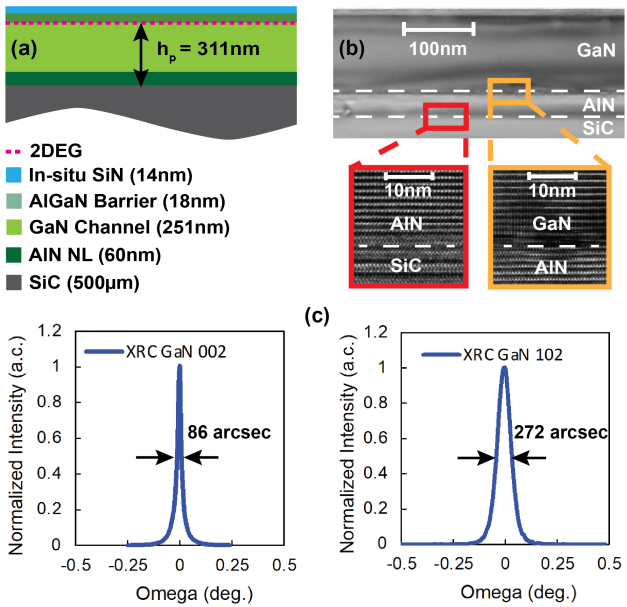

Fig. 1. (a) Cross-sectional schematic of QuanFINE® heterostructure on SiC substrate containing GaN channel on an ultrathin AlN nucleation layer (NL). $h_p$ denotes the combined thickness of the piezoelectric layers i.e., GaN and AlN. (b) TEMs with cross-sections of GaN/AlN/SiC interfaces. (c) XRCs confirming high structural quality of GaN layer.

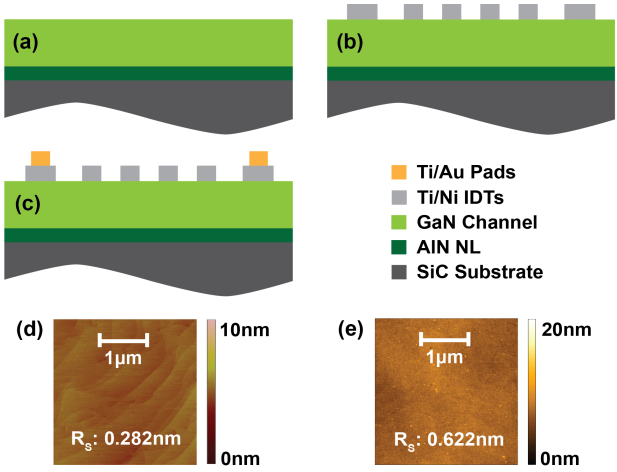

Fig. 3. Fabrication process flow for Sezawa mode devices. (a) Shallow blanket ICP-RIE etch of top SiN and AlGaN layers to eliminate 2DEG. (b) IDT (Ti/Ni) metallization. (c) Pad (Ti/Au) metallization. AFM images show low rms surface roughness ($R_s$) (d) before and (e) after shallow etch.

Finally, Ti(5nm)/Au(245nm) is evaporated and patterned with lift-off using photolithography for the low resistance pads and routings (Fig. 3(c)). Atomic force microscopy (AFM) measurements of the root mean square (rms) roughness of the surface before and after shallow etch (Fig. 3(d, e)) confirm good surface morphology, a necessary property for low propagation loss of the Sezawa mode devices.

Four different types of designs for Sezawa mode devices are studied in this work, including (a) one-port IDTs, (b) one-port resonators confined by shorted metal with one IDT set centered in the cavity, (c) two-port delay lines consisting of two IDT ports separated by an acoustic transmission region, and (d) two-port resonators with metal reflectors forming a resonance cavity and two IDT sets for drive and sense embedded within the cavity. Schematic illustrations alongside optical micrographs of these four are provided in Fig. 4(a-d). Fig. 4(e) shows a scanning electron micrograph (SEM) of the corner of a one-port resonator, with Ti/Ni transducers, reflectors, bus, and Ti/Au pad.

The devices are measured at room temperature in a radio frequency (RF) probe system (Cascade PMC 200) under vacuum. RF input signal of -15 dBm is applied using GSG probes (|Z|, 150 μm pitch) with 50 Ω termination, and scattering parameters (S-parameters) are obtained by a parametric network analyzer (Agilent N5225A). SOLT calibration is performed prior to measurement. The devices under test are then de-embedded from the measured frequency response using open and short structures fabricated on-chip to eliminate electrical parasitics from probe pads and routing to the devices.

## V. Results and Discussions

### A. Equivalent Circuit Models

The cross-field Mason model [34], [35] is modified by adding lumped elements to extract key electromechanical parameters from the measured data for different Sezawa mode devices shown in Fig. 4(a-d). Extracting parameters using the same circuit models provides a common platform to analyze and compare the performances of different SAW designs. Schematic illustrations of the equivalent circuit models for the four different SAW designs are shown in Fig. 5(a-d). For all designs under consideration, a unit cell containing a single IDT finger normalized to the aperture (Fig. 5(e)) is modeled by transmission lines with normalized acoustic impedance for free ($Z_f$) and metalized ($Z_m$) regions of the structure (Fig. 5(f)). The phase angles of the transmission lines ($\phi_f$, $\phi_m$) corresponding to the propagation of the acoustic waves are determined from the finger metallization ratio (m) of the IDT, and the free ($v_f$) and metalized ($v_m$) acoustic velocities calculated for wavelength (λ) [36]. The transformer ratio (η) corresponds to the efficiency of energy conversion from electrical to acoustic domain and vice versa. The capacitance between IDT fingers, including the feedthrough component within the piezoelectric layers, is represented by $C_o$. Finally, electrical losses associated with the piezoelectric transducer are captured in the lumped $R_o$ [37]. To construct the complete model, acoustic ports are cascaded in series, with parallel connection of electrical ports to construct the full N-finger IDT RF port from each IDT pair modeled. The polarity of the transformer swaps for the alternating fingers to model the signal and ground terminals.

As the IDT fingers are electrically shorted at the ends for the reflectors of the resonators, transmission lines representing the acoustic domain without any lumped element are used to denote a unit cell for each finger (Fig. 5(g)). A lossy transmission line of impedance $Z_f$ and phase angle $\phi_p$ is also included in the Mason model for two-port delay lines and resonators to model the acoustic propagation region in the middle (Fig. 5(h)). Additionally, a shunt $C_f - R_f$ branch is used to model electromagnetic feedthrough signal directly between the two ports associated with shorter delay paths [38]. Finally, free propagation away from the structure and its ends are represented by $Z_f$ termination, and electrical resistance and inductance inherent to the pads and transducer fingers are modeled by a series $R_s - L_s$ branch.

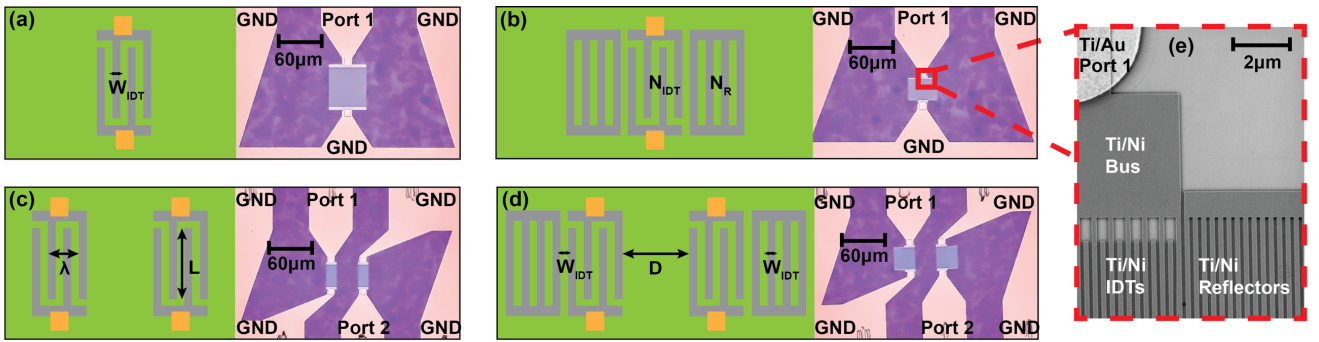

Fig. 4. Schematics of top view (left) and optical micrographs (right) of different SAW designs considered in this study. (a) One-port IDTs. (b) One-port resonator. (c) Two-port delay line. (d) Two-port resonator. (e) SEM of a resonator depicting IDTs, reflectors, pad and bus connectors. Various electrical terminals i.e., RF ports, grounds, and parameters for physical dimensions i.e., unit cell width (λ), IDT width/gap ($W_{IDT}$), port distance (D), device aperture (L), number of IDT ($N_{IDT}$) and reflector ($N_R$) pairs are also illustrated here.

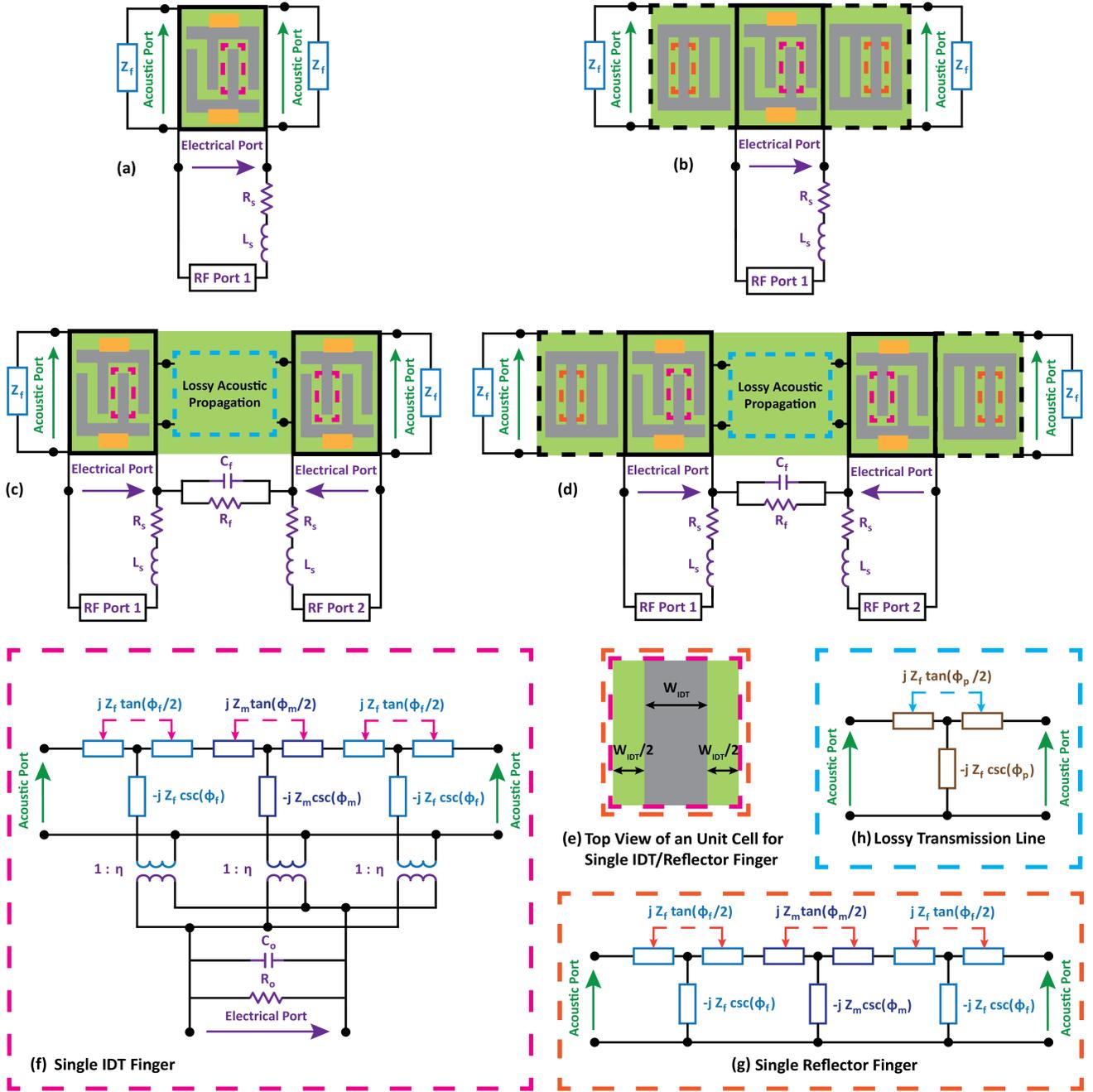

Fig. 5. Schematic illustrations of modified Mason model for one-port (a) IDTs, (b) resonators and two-port (c) delay lines, (d) resonators. (e) A unit cell normalized to the aperture showing top view of a single IDT or reflector finger. (f) An equivalent circuit with transmission lines, lumped elements, and transformers for electromechanical energy conversion representing the unit cell with a single IDT finger for all designs. (g) Transmission lines without lumped elements and electrical port for each unit cell of a shorted reflector for the resonators. (h) A lossy transmission line modeling the acoustic propagation path between the IDTs for two-port delay lines and resonators. Series and parallel connections between acoustic and electrical ports, respectively for full device modeling.

## B. Frequency Responses

Measured frequency spectra for Sezawa mode IDTs, delay lines, and resonators are shown in Fig. 6(a-d). Based on the FEA simulation results, we designed five different Sezawa mode devices for each design category (Fig. 4) with an IDT width/gap between 100-375 nm to investigate a wide operating frequency ranging from 5-14 GHz. The aperture (L) and length of the acoustic transmission path (D) of the devices are selected as an integer multiple of the wavelength (nλ). The IDTs in the reflectors have the same width, gap, and aperture as that of the ports. Measured data from different designs exhibit minimum and maximum resonance frequencies of 5.4 GHz and 14.3 GHz for the IDT width/gap of 375 nm and 100 nm, respectively. So, SAW devices in the QuanFINE® platform exceed resonance frequencies of the state-of-the-art GaN resonators using Rayleigh (8.5 GHz) [24], Sezawa (9.1 GHz) [24], and thickness (8.7 GHz) [39] modes having a comparable IDT resolution. Modified Mason circuit models developed in Fig. 5 for different designs are implemented in Keysight Advanced Design System (ADS) software, and measured data are fitted to extract key performance matrices e.g., electromechanical coupling coefficient ($k^2$), propagation loss ($\alpha$), etc. Only the fundamental Sezawa mode in the frequency response is fitted by the modified Mason model to avoid computation complexities due to a large number of parameters that would arise to incorporate the spurious mode in the model. The zoomed insets of Fig. 6 show ADS fitting with the measurement of the magnitude and phase responses for the IDTs and resonators, and group delay [40] for the delay lines validating our circuit models (Fig. 5) for the 150 nm IDT devices. Parasitic feedthrough for the shorter delay paths and higher-order mode coupled with the fundamental Sezawa mode [12] could explain the mismatch between measured and fitted responses, particularly for the two-port devices in the insets of Fig. 6(c, d). Further analysis in the design space is required for Sezawa devices in the QuanFINE® platform to operate in the single-mode condition, including apodization of the duty cycle of the IDTs, selection of the number of IDT pairs in the ports and reflectors, and adjustment of IDT metal thickness [12].

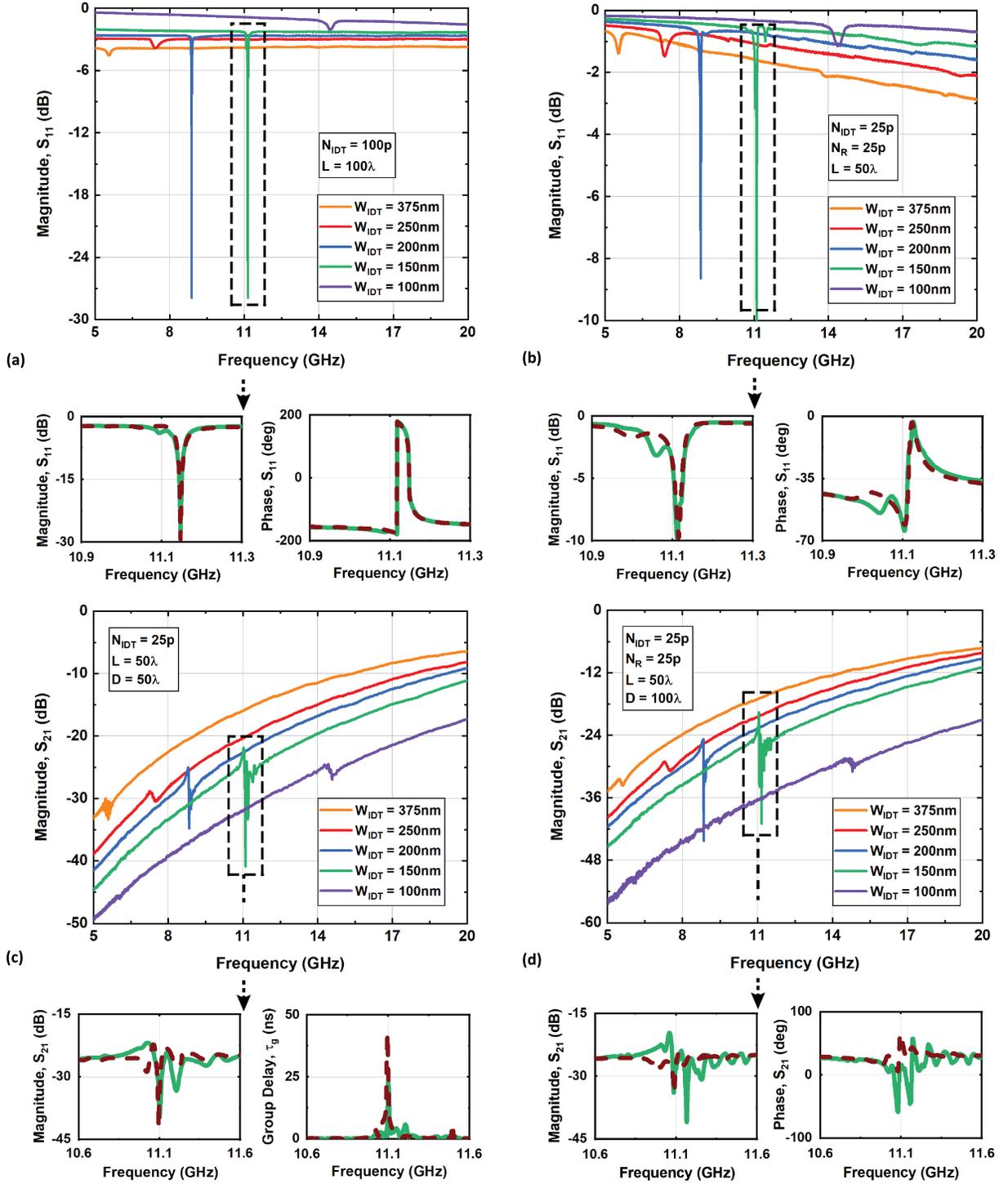

Fig. 6. Measured reflection spectra ($S_{11}$) of one-port (a) IDTs, (b) resonators, and transmission spectra ($S_{21}$) of two-port (c) delay lines, (d) resonators. The legends specify physical device parameters for the devices. Zoomed-in spectra below each plot show measured (solid) and fitted (dash) responses of magnitude and phase/delay for the 150 nm IDT devices.

### C. Phase Velocity

The dependence of phase velocity ($v_p$) on the normalized piezoelectric thickness ($h_p/\lambda$) for the Sezawa mode in the QuanFINE® structure is obtained from simulated and measured resonance frequency ($f_s$) using the following equation [7].

$$v_p = f_s \, \lambda \quad (1)$$

Fig. 7 confirms the high phase velocity of the Sezawa mode in the GaN/SiC platform. A gradual reduction in phase velocity is observed as the acoustic wavelength reduces, presenting a limiting factor for the frequency scaling of Sezawa mode devices for a given lithographic resolution. Experimental results obtained from different SAW designs (Fig. 6) agree well with the 2D FEA simulation.

### D. Electromechanical Coupling Coefficient

Series ($f_s$) and parallel ($f_p$) resonance frequencies (Fig. 2) obtained from the FEA simulation are used to determine the electromechanical coupling ($k^2$) for the Sezawa mode by the following equation [41].

$$k^2 = \frac{\pi^2}{4}\left(1 - \frac{f_s}{f_p}\right) \quad (2)$$

$k^2$ are also calculated from the extracted circuit parameters via fitting the measurements (Fig. 6) using the modified Mason models (Fig. 5) with the following equation [35].

$$k^2 = \frac{\eta^2}{2 f_s C_o Z_\mathrm{f}} \left[\frac{J\{\sin(m\frac{\pi}{2})\}}{J\left(2^{-\frac{1}{2}}\right)}\right]^2 \quad (3)$$

Here, $J$ is the Jacobian elliptic integral [42]. $k^2$ extracted by fitting with the modified Mason models (Fig. 5) follows the trend with the simulation results for all devices, validating the use of these circuit models. Fig. 8 exhibits a maximum $k^2$ value of 0.61% at 11 GHz for a two-port resonator. Considering a unit cell for the IDT fingers normalized to the device aperture (Fig. 5(e)) and lumped resistance ($R_s$) for the distributed losses associated with the IDTs (Fig. 5(a-d)) can explain the deviation in the measurements as compared to the 2D FEA simulations. In all cases, $k^2$ arrives at a maximum value around $h_p/\lambda \sim 0.5$, but decreases gradually with lower or higher values of $h_p/\lambda$. This trend is expected based on the confinement of the Sezawa mode inside the metal IDTs for shorter wavelengths and excessive penetration of the mode into the non-piezoelectric SiC substrate in the case of longer wavelengths. The selection of GaN thickness is necessarily a critical design consideration to achieve a desired frequency range of high-efficiency Sezawa mode devices.

According to theoretical analysis, a maximum $k^2$ of 1.3% and ~2% can be achieved for in-plane and thickness mode devices, respectively, in the GaN platform [11], [43]. Piezoelectric coupling $k^2$ depends on the acoustic mode of vibration, with SAW modes typically demonstrating lower $k^2$ compared to the thickness mode devices [11]. Fig. 9 shows a comparison between the maximum $k^2$ obtained from our

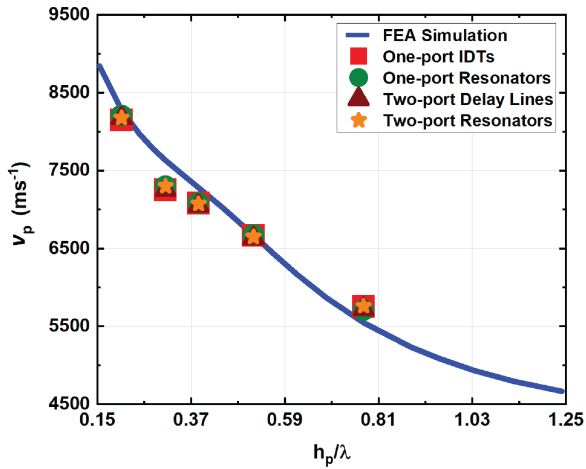

Fig. 7. Simulated (line) and experimental (symbol) results of phase velocity ($v_p$) dependence on the normalized thickness of piezoelectric layers ($h_p/\lambda$).

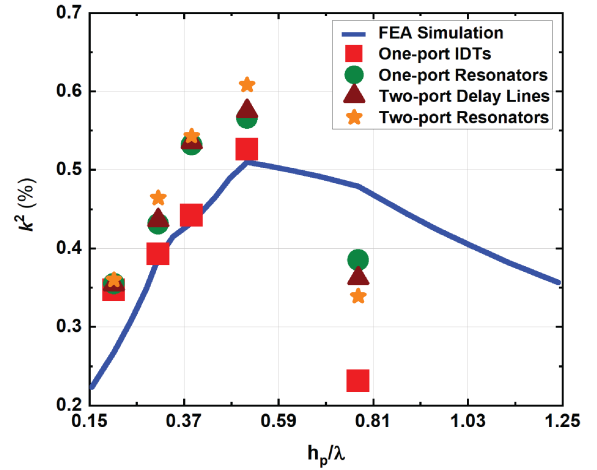

Fig. 8. The dependence of coupling coefficient ($k^2$) obtained from simulated (line) and measured (symbol) data on normalized piezoelectric thickness ($h_p/\lambda$).

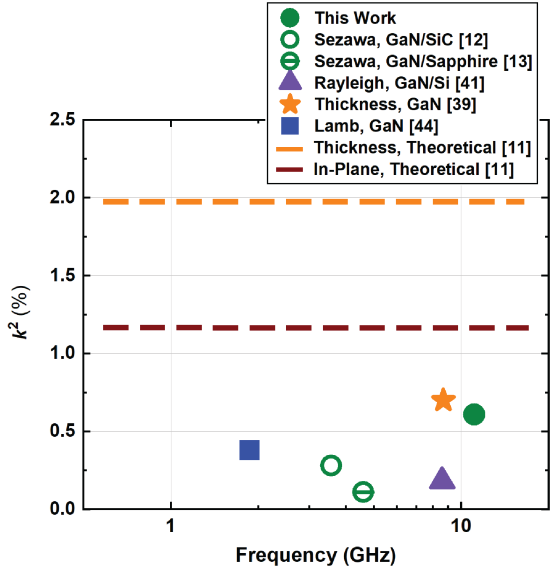

Fig. 9. Coupling coefficient ($k^2$) with the resonance frequencies obtained from our Sezawa mode device compared with other MEMS resonators in GaN MMICs alongside theoretical limits for in-plane and thickness mode in GaN.

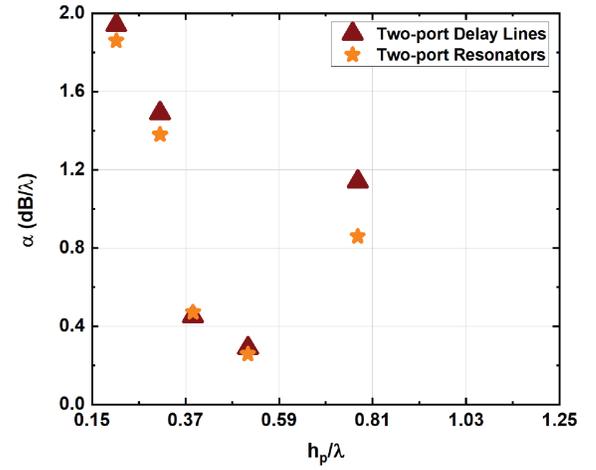

Fig. 10. Propagation loss ($\alpha$) extracted from the measurements by fittings with the normalized thickness of piezoelectric layers ($h_p/\lambda$).

Sezawa mode devices and the state-of-the-art MEMS resonators available in the literature [12], [13], [41], [39], [44] in GaN MMICs. The resonators in this work show 2× higher $k^2$ compared to previously reported Sezawa mode resonator in GaN/SiC, and similar $k^2$ with the thickness mode device in GaN [39]. The relatively high coupling in our devices provides an insight into the defect-free crystalline quality of the piezoelectric layers and effective confinement of acoustic energy for Sezawa mode due to acoustic waveguiding in the QuanFINE® material stack.

Although the maximum $k^2$ achieved from the SAW devices (0.61% at 11 GHz) may seem low for wideband filter application, it can be further extended by incorporating high $k^2$ materials such as GaScN [45] or AlScN [46] in the QuanFINE® structure, or by implementing active transduction mechanisms as in [47]. This work primarily focuses on the feasibility of the platform for MEMS components in the SHF regime and the inclusion of Sc doping in the GaN/AlN stack requires further study. Nonetheless, the devices in this work, with their low propagation loss, can be useful as SAW sensors having high sensitivity due to super-high operating frequencies [7]–[9].

*E. Propagation Loss*

Modified Mason circuit models developed in the previous section (Fig. 5) are used to extract propagation loss ($\alpha$) associated with the acoustic transmission path in two-port delay lines and resonators (Fig. 6(c, d)). Propagation loss is an important performance metric for the two-port devices where acoustic wave attenuates while traveling through a lossy medium between RF ports. Dispersion of the loss with the normalized piezoelectric thickness shown in Fig. 10 reaches a minimum value of 0.26 dB/λ at 11 GHz. For higher $h_p/\lambda$ values, the Sezawa mode is mainly confined at the upper portion of the GaN channel, likely making surface defects a dominant factor for the higher propagation loss. For lower $h_p/\lambda$, the propagating acoustic wave penetrates deeper into the GaN/AlN/SiC interfaces resulting in higher interfacial loss and substrate radiation [48].

*F. Frequency-Quality Factor Product*

A modified Butterworth-Van-Dyke (BVD) model [44], [49] (Fig. 11(a)) is used to extract the frequency-quality factor product ($f \cdot Q_m$) of our one-port resonators (Fig. 6(b)) to benchmark their performance with the state-of-the-art MEMS devices in GaN MMICs. A series $R_m - L_m - C_m$ branch is used to represent the acoustic behavior for series resonance, a shunt $C_o - R_o$ branch denotes capacitance and electrical losses of the piezoelectric transducer, and a $L_s - R_s$ branch models series electrical losses for the IDTs and pads. Multiple motional branches ($R_m - L_m - C_m$) are connected in parallel (Fig. 11(a)) to fit the fundamental Sezawa mode alongside spurious and higher order modes for those devices exhibiting such nonidealities. A sample modified BVD model fit of measured reflection ($S_{11}$) by ADS for both fundamental and spurious modes of a 150 nm IDT device is shown in Fig. 11(b). The unloaded or mechanical quality factor ($Q_m$) and loaded quality factor ($Q_l$) of a particular acoustic mode are obtained using the following equations [44], [50].

$$Q_m = 2\pi f_s \frac{L_m}{R_m} \quad (4)$$

$$Q_l = 2\pi f_s \frac{L_m}{R_m + R_s} \quad (5)$$

Fig. 11(c) shows the measured dependence of $Q_m$ and $Q_l$ on the resonance frequency extracted using the modified BVD model of five one-port resonators (Fig. 6(b)). The figure also includes $Q_m$ and $Q_l$ for higher order spurious modes that exist in the frequency responses of 9 GHz (200 nm IDTs) and 11 GHz (150 nm IDTs) devices (Fig. 6(b)). As the width ($W_{IDT}$) and aperture (L) of the IDTs in the resonators are scaled according to the wavelength (λ), the series resistances for all devices are similar ($R_s$ ~10 Ω). However, comparable motional resistance ($R_m$

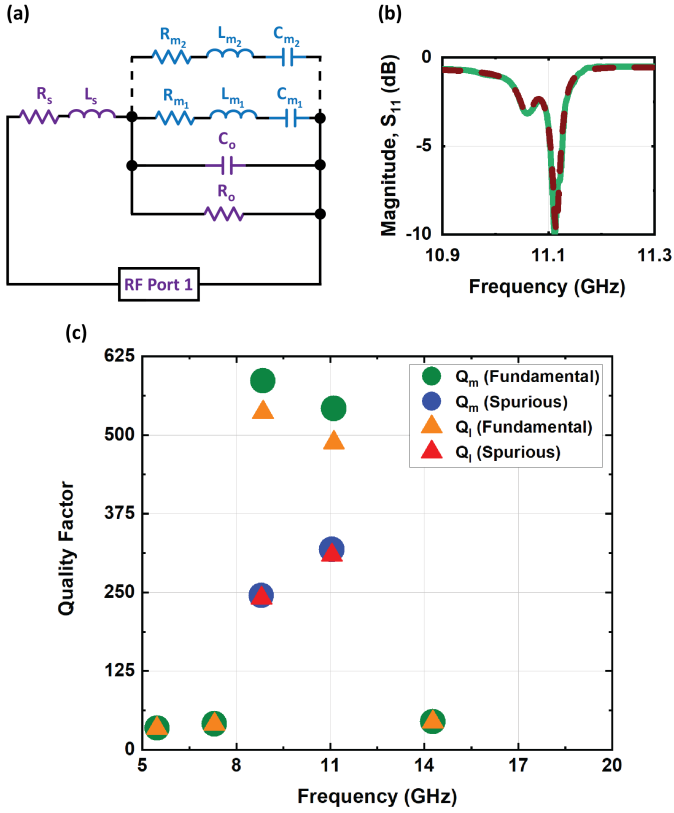

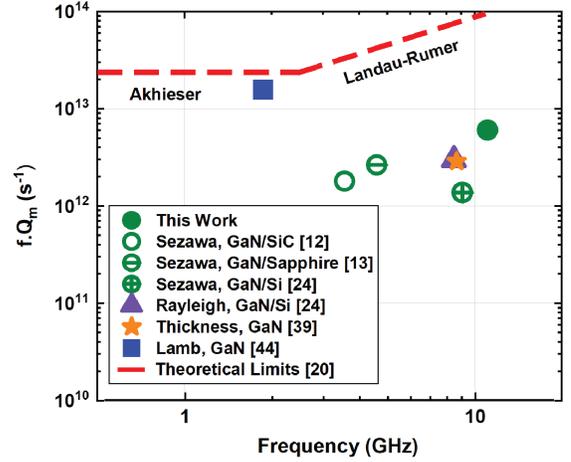

Fig. 12. $f.Q_m$ product with frequencies for the fundamental Sezawa mode of the resonator in this work compared with other MEMS devices in GaN MMICs and the theoretical limits of phonon-phonon scattering for GaN.

Fig. 11. (a) Modified BVD model for one-port resonators showing electrical components (purple), and motional branches (blue) for fundamental and spurious mode. (b) Measured (solid) and fitted (dash) frequency responses for the resonator with 150 nm IDTs. (c) Quality factors ($Q_m$ and $Q_l$) with resonance frequencies for the fundamental Sezawa mode and spurious mode obtained from the measurements by fittings for the resonators.

~90 Ω) with $R_s$ for the fundamental mode causes ~10% lower $Q_l$ than $Q_m$ for 9 GHz and 11 GHz devices. The rest of the resonators exhibit significantly higher $R_m$ (~1 kΩ) than $R_s$ due to the dispersion of the Sezawa mode in the QuanFINE® platform for the layer thickness provided, which results in similar $Q_l$ to $Q_m$.

A maximum $Q_m$ of 542 at 11 GHz is achieved from our resonators as shown in Fig. 11(c). Conversion of the minimum propagation losses [20] (0.26 dB/λ) obtained from the two-port devices in Fig. 10 results in $Q_m$ of 104, which is 5× lower compared to that of the resonators. The lower $Q_m$ of the two-port transmission structures points to scattering at the interface of the IDT and metal-free propagation region. In the resonator, the periodic metallization continues seamlessly into the reflectors, preventing this radiative loss mechanism. This observation is supported by the FEA simulation available in the literature [51].

The $f.Q_m$ product for each resonator is calculated by multiplying the resonance frequency ($f_s$) of a mode by its extracted $Q_m$. Fig. 12 compares the $f.Q_m$ product obtained for the fundamental mode of our resonators against Sezawa mode devices on different substrates [12], [13], [24] as well as other electromechanical modes [24], [39], [44] in GaN MMICs, alongside the fundamental limits obtained from the theoretical calculation for the phonon-phonon scattering in GaN [20]. A maximum $Q_m$ of 542 results in $f.Q_m$ product of $6×10^{12}$ s$^{-1}$ at 11 GHz, which is 6× higher compared to the resonators previously reported as a record high frequency (9.1 GHz) in the literature for Sezawa GaN [24]. Scattering of acoustic waves due to the phonon-electron interaction in the UID GaN layer and mass loading for the metal IDTs can result in $f.Q_m$ product lower than the theoretical limits [20]. Although Lamb mode resonators at 1.87 GHz in Ref. [44] showed $f.Q_m$ products close to the fundamental limits, narrow tether design, and non-linearity associated with the thin membrane of the released devices make them challenging to scale to the SHF regime.

## VI. Conclusion

The QuanFINE® platform is shown in this work to enable frequency scaling beyond the state-of-the-art MEMS devices in GaN MMICs with simple fabrication. Extracted phase velocity and electromechanical coupling from modified Mason model fittings for different SAW designs show the dispersion of the Sezawa mode in the structure and close agreement with FEA simulation. The resonators exhibit low propagation loss and high frequency-quality factor product towards the fundamental limits at record high frequency in GaN MMICs. This technology with AlGaN/GaN heterostructure and high 2DEG density provide a platform for low-cost and high-performance reconfigurable MEMS components in GaN MMICs for programmable ad-hoc radios in the super high frequency regime [52], [53], and opening doors for high-performance SAW sensors, monolithically integrated with peripheral active circuitries in GaN MMICs for harsh environments [54], [55].

## Acknowledgment

Microfabrication and characterization of the SAW devices were performed at the Birck Nanotechnology Center at Purdue University. The authors are thankful to Jackson Anderson from

HybridMEMS Research Group for his support during the measurements.

BIOGRAPHIES

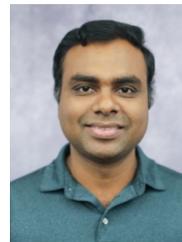

**Imtiaz Ahmed** (Graduate Student Member, IEEE) received B.Sc. (Hons.) and M.Sc. degrees in Electrical and Electronic Engineering (EEE) from Bangladesh University of Engineering and Technology (BUET), Dhaka, Bangladesh in 2012 and 2015, respectively. He is currently pursuing a Ph.D. degree in the Elmore Family School of Electrical and Computer Engineering at Purdue University, West Lafayette, IN, USA. He is also serving as a Graduate Research Assistant in Hybrid MEMS Lab at Purdue University. He worked as an Assistant Professor in the Department of EEE at BUET from 2016 to 2017 and currently he is on study leave to pursue his Ph.D. at Purdue University. His research interests include the design, modeling, and fabrication of monolithically integrated MEMS resonators and non-reciprocal devices in GaN MMIC technology. He is also interested in MEMS-based ultrasonic transducers.

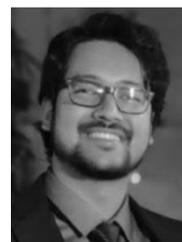

**Udit Rawat** (Graduate Student Member, IEEE) received B.Tech. degree (Hons.) from Uttarakhand Technical University, Sudhowala, India, in 2012, and M.S. (Research) degree from IIT Madras, Chennai, India, in 2015. He is currently pursuing a Ph.D. degree with the Elmore Family School of Electrical and Computer Engineering, Purdue University, West Lafayette, IN, USA. From 2015 to 2017, he worked as a Digital Design Engineer at

Intel Corporation, Bengaluru, India. He is currently a Research Assistant with the Hybrid MEMS Research Group at Purdue University. His research interests include the design, fabrication, and modeling of RF MEMS resonators as well as their integration into standard CMOS technology. He is also interested in the design of unreleased CMOS-integrated gyroscopes and other physical sensors along with their interface circuit design.

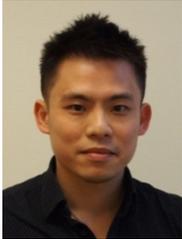
**Jr-Tai Chen** received an M.Sc. degree in Electro-Optical Engineering from Tatung University, Taipei, Taiwan in 2007, and a Ph.D. degree in Material Engineering from Linköping University, Linköping, Sweden in 2015. In 2014, he co-founded SweGaN AB, Linköping, for manufacturing high-quality GaN-based epi wafers, where he is currently a Chief Technology Officer.

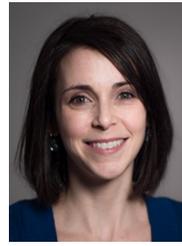
**Dana Weinstein** (Senior Member, IEEE) received a B.A. degree in physics and astrophysics from the University of California, Berkeley, CA, USA in 2004 and a Ph.D. degree in applied physics from Cornell University, Ithaca, NY, USA in 2009, working on multi-GHz MEMS. She is currently an Professor at the Elmore Family School of Electrical and Computer Engineering, and the Associate Dean of Graduate Education, College of Engineering at Purdue University, IN, USA. Prior to joining here, she worked in the Department of Electrical Engineering and Computer Science at Massachusetts Institute of Technology (MIT), Cambridge, MA, USA as an Assistant Professor and served as an Associate Professor there between 2013 and 2015. She is the recipient of the Transducers Early Career Award, the NSF Career Award, the DARPA Young Faculty Award, the Intel Early Career Award, and the IEEE IEDM Roger A. Haken Best Paper Award. Her current research focuses on hybrid MEMS-IC devices for low-power wireless communication, microprocessor clocking, and sensing applications. In particular, she is working to harness the benefits of acoustic vibrations to enhance the performance of next-generation electron devices.